\documentclass{article}

\usepackage[english]{babel}
\usepackage{amsmath}
\usepackage[letterpaper,top=2cm,bottom=2cm,left=3cm,right=3cm,marginparwidth=1.75cm]{geometry}

\usepackage{graphicx} 
\usepackage{caption}
\usepackage{subcaption}
\usepackage{authblk}
\usepackage{hyperref}

\title{Architecture-dependent Effects of Trade-offs on Evolutionary Navigability and Epistasis}
\author{Nandita Chaturvedi
\thanks{cnandita@ncbs.res.in}}
\affil[1]{National Institute of Advanced Studies, Bangalore, India}

\begin{document}
\date{}
\maketitle

\begin{abstract}
The structure of the genotype-phenotype map plays a central role in determining how populations move through fitness landscapes, yet less is known about how this structure interacts with environmental constraints and fitness trade-offs. Here, we study evolution on a tunable multilayer genotype-phenotype map in which binary genotypes are transformed through successive feed-forward layers before being evaluated by phenotype-level
fitness functions. We explicitly model the phenotype to fitness transformation in order to compare an emergent trade-off fitness with a no-trade-off control. We explore different map architecture as models of more and less complex genotype-phenotype relationships by tuning the genotype length and number of layers. We find that ruggedness and navigability have a complex, architecture-dependent relationship. Landscapes with few local maxima can still have low navigability for wide and shallow maps, and navigability is high even at mean values of ruggedness for wide and deep maps. Trade-offs increase navigability in wide and shallow maps, while their effect is muted in other architectures. The effect of trade-offs on epistasis depends on evolutionary history. While trade-offs change which mutational neighborhoods are sampled by evolution, the background degree of epistatic interactions is set by map architecture. Finally, we use our framework to investigate micro-evolution of high-fitness populations and show that temporal covariance in allele-frequency changes can arise from the internal structure of a rugged genotype-phenotype map, producing signatures that resemble short-term alternating selection.
\end{abstract}

\section{Introduction}

How phenotypes emerge from underlying genotypes and epigenetic processes remains a central question in developmental and evolutionary biology \cite{manrubia2021genotypes}. Environmental selection and evolutionary pressures interact directly with the phenotype and yet rely upon mutations at the genotypic level to carry out evolutionary searches. This two-fold evolutionary process can be captured by the notion of a genotype-to-phenotype map, and has wide ranging implications for evolution. For one, the direction and efficiency of the evolutionary search is determined in large part by the structure of such a genotype-phenotype map \cite{pigliucci2010genotype}. 

Here we study a rugged fitness landscape that emerges from a multi-layer genotype to phenotype map. We explicitly model how environmental constraints enter through the mapping from phenotype to scalar fitness \cite{chevin2022using}. We explore different architectures of the genotype to phenotype map and consider varying degrees of complexity. The depth of our map, $L$, characterizes the degree to which phenotype is mediated by successive transformations between genotype and fitness. Past studies have largely considered single layer transformations or a combination of a few separate layers. Others have looked at the effect of hierarchy through a fixed number of intermediate levels \cite{arias2014toylife}. We ask how different quantitative measures such as navigability and ruggedness of the map change when competing fitness effects are considered rather than a single fitness objective. Through our map we are able to probe the changes that come with varying depth alongside varying genotype size, $N$. Low-depth maps provide a model for relatively direct genotype–phenotype relationships, while high-depth maps represent traits whose effects are filtered through multiple regulatory, developmental, or physiological stages. In this sense, varying $L$ allows us to ask how the consequences of trade-offs change with increasing genotype–phenotype complexity. The explicit modeling of the phenotype level and fitness function allows us to compare the effect of a fitness trade-off with a control no-trade-off fitness objective for varying map complexities. 

There have been attempts to characterize the structure of complex genotype-phenotype maps, both empirically derived as well as those derived from theoretical considerations \cite{wagner2011pleiotropic, greenbury2022structure, ahnert2017structural, greenbury2015organization, fortuna2017genotype}. The ability of evolution to find globally optimum genotypes has been quantified using measures such as navigability and evolvability \cite{papkou2023rugged, manrubia2017distribution}, to which ruggedness is related. Initial studies suggested that more rugged fitness landscapes are less navigable \cite{kauffman1989nk, kauffman1987towards}, while more recent work has shown that high dimensional genotypic spaces can make even rugged landscapes navigable \cite{papkou2023rugged}. The trade-off between robustness and navigability has also received much attention \cite{ciliberti2007innovation, wagner2008robustness, weiss2018phenotypes}.

Complex genotype-phenotype maps show a high degree of epistasis \cite{szendro2013quantitative, de2014empirical}. This means that the strength and even direction of mutational effects can vary with the background fitness of the organism due to interactions between genetic sites \cite{barker2015dynamic}. Mutations are not necessarily additive and the fitness landscape can have multiple peaks corresponding to the same fitness. Epistasis has been linked to increased ruggedness, and lower navigability \cite{franke2011evolutionary, de2014empirical, poelwijk2011reciprocal, romero2009exploring}. In theoretical work, there are also examples of epistatic interactions speeding up adaptation rather than retarding it \cite{draghi2013selection, greenbury2022structure}. The relationship of environmental trade-offs to epistatic interactions, however, remains relatively unexplored \cite{hall2019environment, bank2022epistasis, de2013environmental, li2018multi}. 

Populations rarely evolve under single fitness goals, but rather try to maximize multiple fitness objectives that can show trade-offs. Microbial systems such as yeast and \textit{E. coli}, for example, exhibit trade-offs between survival in stresses such as from drug exposure or temperature variation on the one hand, and growth rate on the other \cite{zakrzewska2011genome, chaturvedi2026adaptation, li2018hidden, vincent2013fitness, bennett1993evolutionary}. The effective structure of the genotype-to-phenotype map depends not only on epistatic interactions, but also on constraints coming from such fitness trade-offs \cite{el2014genotype, chaturvedi2026adaptation}. Studies on the effect of trade-offs on evolution, such as those on adaptation to varying environments typically do not consider their interaction with complex genotype-phenotype map structure and its epistatic interactions \cite{shoval2012evolutionary, xue2019environment, chaturvedi2025evolutionary}.

We study the effect of environmentally imposed fitness trade-offs on commonly studied characteristics such as ruggedness, navigability and epistasis across different map architectures and complexity. We ask whether navigability and ruggedness have a predictable relationship across map architectures, and whether trade-offs significantly change these characteristics. Lastly, we consider a biological application of our framework. We ask whether neutral evolution of a population of high fitness phenotypes is significantly different with and without fitness considerations that show a trade-off. Recent work on a 10 year survey of the water flea \textit{Daphnia pulex} \cite{lynch2024genome} has put forward the hypothesis that in a high fitness population alternating changes in correlations of minor allele frequency changes across generations points to alternating microselection. We find that a similar pattern also emerges from the presence of multiple near equivalent fitness peaks because of epistasis. 

\section{Methods}
\subsection{Model: Multilayer Genotype-Phenotype Map with Emergent Trade-Offs}

We study evolution on a multilayer genotype-to-phenotype map. A genotype is represented as a binary vector

\begin{equation}
g = (g_1,g_2,\ldots,g_n),
\end{equation}

where each $(g_i \in {0,1})$. This genotype forms the zeroth layer of a feed-forward developmental map. The map contains $L$ layers above the genotype, with each layer containing $N$ binary nodes. Thus, the full developmental state consists of layers

\begin{equation}
\ell = 0,1,\ldots,L,
\end{equation}

where $\ell=0$ is the genotype and $\ell=L$ is the phenotype layer.

The connections between consecutive layers are chosen randomly and then held fixed for a given map. Each node in layer $\ell+1$ receives $k=3$ inputs from nodes in layer $\ell$. Inputs are sampled independently from the previous layer, so the same lower-layer node may feed into multiple upper-layer nodes, and some lower-layer nodes may not be used. Each edge is assigned a sign. With probability $q$, an edge is negated; otherwise it is transmitted without negation. Thus, $q$ controls the fraction of inhibitory or inverted inputs in the map. We fix $q$ to be $0.3$ for the purpose of our study.

The state of each node in the next layer is determined by a majority rule applied to its three signed inputs. If the signed inputs to node $j$ in layer $\ell+1$ are $x_1,x_2,x_3 \in {0,1}$, then the output is

\begin{equation}
x_j^{(\ell+1)} =
\begin{cases}
1, & x_1+x_2+x_3 \geq 2, \\
0,  & x_1+x_2+x_3 < 2.
\end{cases}
\end{equation}

A negated edge changes an input value $x$ to $(1-x)$ before the majority rule is applied. Therefore, once the random wiring and edge signs are fixed, the map deterministically sends every genotype $g$ to a final-layer phenotype vector.

We assume this map to be an abstract model of different complex epigenetic and developmental factors that map the genotype to a phenotype. Different realizations of the random wiring define different genotype-to-phenotype maps.

\subsection{Fitness}
From the final phenotype layer, we construct quantitative phenotypic variables by grouping subsets of final-layer nodes into phenotype bins. If $S_a$ is a subset of phenotype-layer nodes, we define the corresponding phenotypic performance variable as

\begin{equation}
F_a(g) = \sum_{i \in S_a} \phi_i(g).
\end{equation}

Thus, each $F_a$ measures the number of active phenotype nodes within a particular subset of the final layer. For our study we will consider different phenotypic bins to be of length $S_a=N/32$.

To construct a trade-off between two phenotypic objectives, we first sample a collection of $2^{N-1}$ random genotypes and evaluate the final-layer phenotype nodes across this sample. We then identify phenotype-layer nodes or groups of nodes whose values are negatively correlated across sampled genotypes. These negatively correlated nodes are assigned to two distinct phenotype bins, defining two performance variables $F_1(g)$ and $F_2(g)$. Because these variables are constructed from negatively correlated components of the phenotype layer, simultaneous improvement in both is constrained by the structure of the map. They therefore represent an emergent developmental trade-off.

As a control, we also define a third performance variable $F_3(g)$, constructed from a randomly chosen subset of phenotype-layer nodes. This variable is not deliberately chosen to be negatively correlated with $F_1$. We use $F_3$ to define a no-trade-off comparison fitness.

We then compare two scalar fitness functions. The trade-off fitness with the negatively correlated $F_1$ and $F_2$ is

\begin{equation}
F_{\mathrm{TO}}(g)= p_E F_1(g) + (1-p_E)F_2(g), \label{eq:fto}
\end{equation}

where $p_E$ controls the relative weighting of the two competing phenotypic objectives. The trade-off is thus emergent through the network architecture and is then imposed using this scalar fitness function. The no-trade-off control fitness with the uncorrelated $F_1$ and $F_3$ is

\begin{equation}
F_{\mathrm{NTO}}(g)=p_E F_1(g) + (1-p_E)F_3(g).\label{eq:fnto}
\end{equation}

The comparison between $F_{\mathrm{TO}}$ and $F_{\mathrm{NTO}}$ allows us to isolate the effect of an imposed phenotypic trade-off while keeping the overall form of the fitness function fixed.

\begin{figure}%
    \centering    
  {\includegraphics[width=\linewidth]{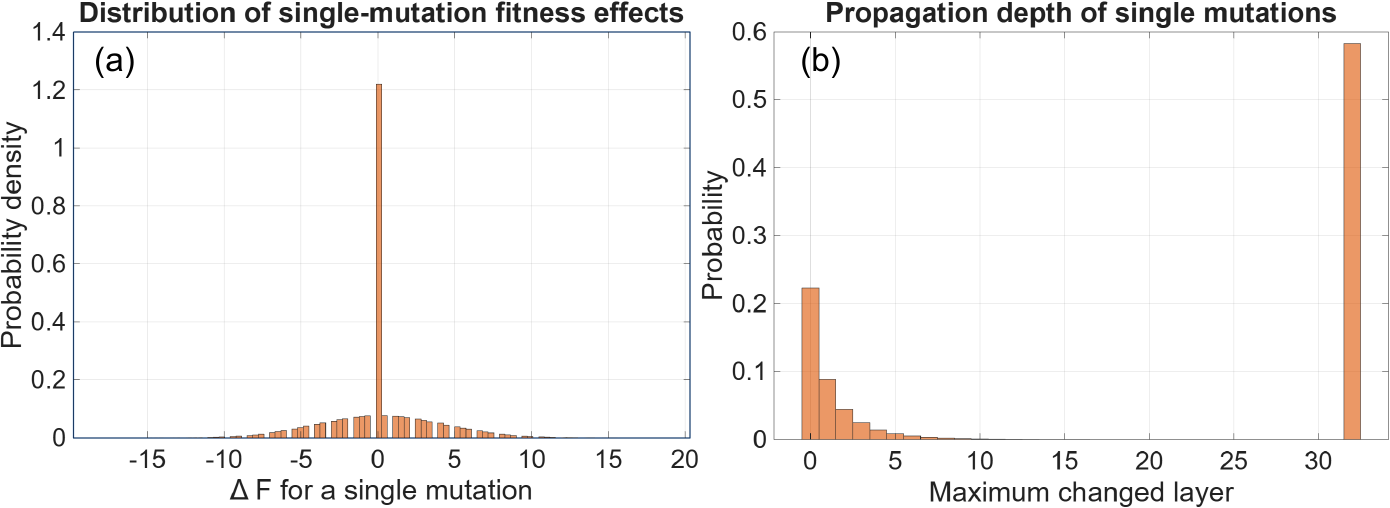}}\\%

   \caption{(a) Probability density of the distribution of fitness effects for a single mutation, averged over 20 maps and $1000$ randomly chosen genotypes for each map. As can be seen, most mutations are neutral (b) Histogram of the propagation depth for a single mutation. Almost half of all mutations are dampened by epistatic effects before they cause a change in the phenotypic layer. The histogram is the average of $20$ maps. \label{fig:distoffit} }
\end{figure}

 Figure \ref{fig:distoffit} shows the probability distribution of fitness effects for a single mutation without a trade-off for $1000$ different genotypes and $20$ different maps with $N=4096$ and $L=16$. The corresponding distribution with a trade-off looks almost identical. The distribution is the average over $20$ different maps, and $1000$ random genotypes tested for each map. The distribution shows that a majority of mutations are neutral. Thus, our network architecture and fitness definition captures this significant property of biological systems. Figure \ref{fig:distoffit} is a histogram of the propagation depth for single mutations, or the maximum layer that changes when the mutation is made. Over $40\%$ mutations fail to reach the top layer, and epistatic effects dampen their influence before the phenotypic layer. This shows the effectiveness of our multilayer architecture in reproducing the epistasis in biological systems. We will study the degree and quality of epistasis further in section \ref{sec_epi}. Lastly, the distribution of fitness effects is symmetric about zero. 

\subsection{Evolutionary Search and Pareto Front}

We run Monte-Carlo simulations to reach the global fitness optimum for $F_{TO}$ in our emergent fitness landscape. We start with $50$ random low fitness genotypes and allow them to randomly mutate at one genotypic site every step in the simulation. For each mutation, a new fitness is calculated and compared to the old fitness. Following \cite{nallaperuma2019analysis}, the mutation is then accepted with a probability given by

\begin{equation}
    p_a=(1-e^{-\frac{2(F_{new}-F)}{kT}})/(1-e^{-\frac{2Npop(F_{new}-F)}{kT}})
\end{equation}

where $F_{new}$ denotes the fitness of the genotype after the mutation and $F$ is its fitness before it. $N_{pop}$ represents the population size we are considering, and $T$ is an effective temperature. This probability allows the search to sometimes accept deleterious mutations, thus allowing the evolutionary search the possibility of exiting from local fitness maxima. The ease with which deleterious mutations can be accepted increases with increasing effective temperature and with population size. We set $kT=0.4$ for our simulations.

\begin{figure}%
    \centering    
  {\includegraphics[width=0.5\linewidth]{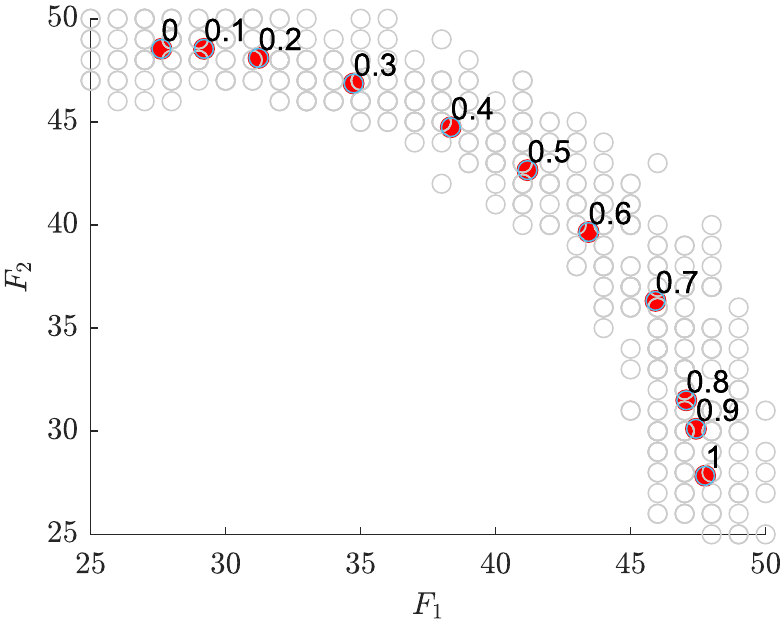}}\\%

    \caption{Pareto front with $F_1$ on the $x$ axis, and $F_2$ on the $y$ axis. Each red point represents the mean fitness results of an evolutionary search for $10$ different genotypes and $10$ trials for each, with the value of $p_E$ for each labeled next to it. The gray points show the results of each of these individual searches. As can be seen, the Pareto front has a finite width.   \label{fig_pareto} }
\end{figure}

Depending on the value of $p_E$, the optimum phenotype lies between the two archetypes that correspond to $F_1$ and $F_2$. For $p_E=0$, we expect to maximize $F_2$, and for $p_E=1$ we expect to maximize $F_1$. For all intermediate values, we find that the fitness $F$ is maximized for varying weights of $F_1$ and $F_2$. This is shown in Figure \ref{fig_pareto}. 

The curve for all $p_E$ represents the Pareto front for the competing maximization goals of $F_1$ and $F_2$. It also corresponds to the fitness set from Richard Levins' work on optimal phenotypes in heterogeneous environments \cite{levins1963theory}. Our Pareto front has a finite width, reflecting the fact that the evolutionary search is unable always to find the global fitness optimum, and can be stuck in a local optimum that still has a high total fitness. This width is dependent on the temperature we set for the  evolutionary search and reflects the ruggedness of the genotype-phenotype map. For our exploration of ruggedness, navigability and epistasis, we will use $p_E=0.5$ so that each fitness consideration is equally important.

\subsection{Navigability and Ruggedness}

To quantify the structure of accessible evolutionary routes, we measured navigability and ruggedness for each genotype--phenotype map under both the trade-off fitness function and the no-trade-off control. For each map, we first estimated the distribution of fitness values by evaluating a set of randomly
sampled genotypes. This sampled fitness distribution was used to define a low-fitness starting range and a high-fitness target range for each map and fitness condition. For every map, we then generated 100 start--target fitness pairs, with starting fitnesses sampled near the lower end of the fitness distribution and target fitnesses sampled near the upper end. We consider different network architectures and repeated the process across six width-to-depth ratios, with $(N,L)=\{(1024,64),(1024,32),(2048,32),(4096,32),(4096,16), (4096,8)\}$. Thus, we investigated different width to depth ratios,
\begin{equation}
    N/L = (16,\;32,\;64,\;128,\;256,\;512) .
\end{equation}

Following \cite{papkou2023rugged}, we define navigability as the probability that a low-fitness starting genotype can reach a high-fitness target region by such a strictly uphill path \cite{weinreich2006darwinian, franke2011evolutionary}. Ruggedness is estimated as the fraction of sampled genotypes that are local maxima, meaning that none of their one-mutant neighbors has higher fitness.

For computational tractability, the accessible-path search was restricted to a fixed $N/2^7$-site mutable subspace of the full genotype. A starting genotype was chosen by randomly sampling genotypes until one was found whose fitness lay within a fixed tolerance of the chosen starting fitness. From this starting genotype, we performed a breadth-first search on the directed graph of accessible one-mutant moves. A one-mutant move was considered accessible only if it produced a strict increase in the scalar fitness. Thus, the search only explored paths along which fitness increased monotonically at every mutational step. A target was counted as reached only when it was reached through such an uphill move. One-mutant neighbors were generated by flipping a single mutable bit, and previously visited genotypes were stored to avoid repeated exploration.

For a given start-target pair, navigability was assigned a value of one if the breadth-first search found at least one strictly uphill path from the starting genotype to the target fitness band, and zero if no such path was found within the search cutoff. Cases in which no suitable starting genotype could be found were excluded from the average. The navigability of a map was then defined as the mean success probability over the 100 start-target pairs.

Ruggedness was measured during the same accessible-path search. Each explored genotype was classified as a local maximum if none of its one-mutant neighbours within the mutable subspace had higher fitness. The ruggedness of a map was then calculated as the fraction of explored genotypes that were local maxima. Thus, ruggedness is a local property of the accessible graph, whereas navigability measures whether locally accessible steps can be connected into a successful multi-step route. Both these measures should give us a sense of the efficiency of the evolutionary search in our constructed landscape, and allow us to investigate how the introduction of fitness trade-offs changes this efficiency.

We repeated this calculation for both scalar fitness functions, with and without trade-offs as defined in equations \ref{eq:fto} and \ref{eq:fnto}. The same procedure was applied across 20 independently generated maps for each architecture. Note that $N/L=16, 64, 256$ all have the same total number of nodes yet we will show they exhibit different navigability and ruggedness trends in Figure \ref{fig_nav}.

To identify which aspects of the landscape were responsible for the observed changes in navigability, we computed several additional diagnostics for each genotype-phenotype map and fitness condition. These diagnostics were measured separately for the trade-off fitness function \(F_{\mathrm{TO}}\) and the
no-trade-off control \(F_{\mathrm{NTO}}\), and then compared using paired differences within the same map.

For each diagnostic \(X\), we computed the paired trade-off effect 
\begin{equation}
\Delta X = X_{\mathrm{TO}} - X_{\mathrm{NTO}}
\label{eq:diff}
\end{equation}

for each independently generated map. Negative values therefore indicate that the quantity is lower under the trade-off fitness function than under the no-trade-off control, while positive values indicate that it is higher under the trade-off. The plotted points in Figure \ref{fig:diag} show the mean of \(\Delta X\) across maps for each architecture, and error bars denote the standard error across maps. 

First, we plot the change in the ruggedness or fraction of local maxima because of the introduction of the trade-off. Second, we measured the mean beneficial step size. For each map and fitness condition, we sampled genotypes and computed the fitness effects of all one-mutant neighbors in the mutable subspace. We then averaged only over the positive fitness effects, $\left\langle \Delta F \mid \Delta F > 0 \right\rangle$. This quantity measures the typical size of an available uphill move.

Third, we recorded the length of successful accessible paths. In the navigability calculation, a breadth-first search was performed on the directed graph of strictly fitness-increasing one-mutant moves. For every successful search, the accessible path length was defined as the number of one-mutant steps in the first path that reached the target fitness band. Because breadth-first searches explore paths in order of increasing length, this corresponds to the shortest strictly uphill path found within the mutable subspace. The mean path length was computed only over successful searches.

Finally, we measured search effort by recording the number of distinct genotypes visited during the search. Unlike path length, this quantity was recorded for both successful and unsuccessful searches, and measures how much of the accessible neighborhood had to be explored before either reaching the target or terminating at the search cutoff.

These diagnostics separate local landscape structure from global path structure; the fraction of local maxima and beneficial step size describe local mutational neighborhoods, whereas accessible path length and visited count describe how those local neighborhoods are connected into evolutionary routes.

\subsection{Epistasis}

To quantify how the genotype--phenotype map shapes background-dependent mutation effects, we performed an epistasis sweep across our set of multilayer architectures used in the navigability analysis. For each architecture, we used independently generated maps and restricted the mutational analysis to a fixed set of $N/2^7$ mutable sites. 

For each map, we considered three classes of genetic backgrounds: randomly sampled genotypes, genotypes evolved under the trade-off objective (TO population), and genotypes evolved under the no-trade-off objective (NTO population). At each background, we evaluated the fitness effect of single mutations at the fixed mutable sites under both fitness definitions.

For a mutation at site \(i\), the single-mutant fitness effect was defined as a function of the final layer,
\[
\Delta F_i(g) = F(g_i') - F(g),
\]
where \(g_i'\) denotes the genotype obtained by flipping site \(i\) in the background genotype \(g\). We then asked how much \(\Delta F_i(g)\) varied across genetic backgrounds. This provides an epistasis-like measure of background-dependence: if the effect of a mutation depends strongly on the rest of the genotype, then that site is strongly epistatic.

We summarized this background-dependence using three quantities. First, for each site we computed the standard deviation of \(\Delta F_i\) across genetic backgrounds, and then averaged this quantity over sites. This measures the typical background-dependence of mutational effects, $\left\langle \mathrm{std}_g \left(\Delta F_i(g)\right) \right\rangle_i$. 
Second, for each site we computed the range of \(\Delta F_i\) across backgrounds, again averaging over sites,
\[
\left\langle
\max_g \Delta F_i(g) - \min_g \Delta F_i(g)
\right\rangle_i .
\]
This measures the extreme background-dependence of mutational effects. 

The standard deviation and range of \(\Delta F\) are complementary measures. The standard deviation measures typical variation in the effect of a mutation across genetic backgrounds, whereas the range is sensitive to the most extreme backgrounds observed. Thus, if both quantities show the same trend, the change in background-dependence is distributed broadly across the sampled backgrounds. If the range changes more strongly than the standard deviation, the effect may be driven by rare backgrounds with unusually large or unusually small mutation effects.

To compare the trade-off and no-trade-off landscapes directly, we once again computed paired differences of these quantities of the form in equation \ref{eq:diff}. Negative values therefore indicate that the trade-off
objective has lower background-dependence than the no-trade-off objective, whereas positive values indicate that the trade-off objective has higher background-dependence.

\subsubsection{Microevolution in a High Fitness Population}

Ecological studies often collect data from populations that are undergoing low levels of selection, either from intrapopulation variation and population density effects, or from short term changes in the environment. We consider a recent 10 year population-genomic survey of the microcrustacean \textit{Daphnia pulex} \cite{lynch2024genome}. The study finds that microselection from a varying environment could be the cause for standing genetic variation in the population.

In order to examine the results from this study, we prepare a population of $1000$ individual genotypes. Using a map architecture with $N=4096$ and $L=16$, we allow the population to evolve under three paradigms: with a fixed fitness equivalent to $F_1$, with a fitness that varied every generation between $F_1$ and $F_2$, and with a fitness that varied every generation between $F_1$ and $F_3$. In other words, we subjected the population to constant selection, to varying selection between two fitness measures that did not exhibit a trade-off, and two that exhibited a fitness trade-off. Note that cases with alternating selection every generation can be qualitatively compared to selection under a total fitness $0.5(F_1+F_2)$ or $0.5(F_1+F_3)$ respectively. We allow the population to reach high fitness under each paradigm.

Following \cite{lynch2024genome}, we calculate the covariance in minor allele frequency changes for our population. Using the methodology of  \cite{buffalo2019linked}, we quantified temporal structure in allele-frequency change by computing the genome-wide covariance of changes in minor-allele frequency. For each genetic site which shows variation in the population, \(i\), we recorded the minor allele frequency \(p_i(t)\) at generation \(t\), and defined the allele-frequency change over one sampling interval as
\[
\Delta p_i(t) = p_i(t+\Delta t)-p_i(t).
\]
We then computed the covariance between allele-frequency changes separated by a time lag \(\tau\) across loci:
\[
C(t,\tau)
=
\frac{1}{M}
\sum_{i=1}^{M}
\frac{
\left(\Delta p_i(t)-\overline{\Delta p(t)}\right)
\left(\Delta p_i(t+\tau)-\overline{\Delta p(t+\tau)}\right)
}{
\overline{p_i}(1-\overline{p_i})
},
\]
where \(M\) is the number of segregating sites included in the calculation, \(\overline{\Delta p(t)}\) denotes the mean allele-frequency change across sites during interval \(t\), and \(\overline{p_i}\) is the mean frequency of the minor allele at site \(i\) over the relevant time points. We normalized by $\overline{p_i}(1-\overline{p_i})$ so that covariance values were not dominated by loci with intermediate allele frequencies, which have larger expected allele-frequency fluctuations. 

For each lag \(\tau\), we averaged \(C(t,\tau)\) over all pairs of time intervals separated by that lag,
\[
C(\tau)=\left\langle C(t,\tau)\right\rangle_t .
\]
Positive values of the covariance indicate that allele-frequency changes tend to persist in the same direction across time intervals, whereas negative values indicate reversals in the direction of allele-frequency change. The sign of covariance values has been interpreted in the literature to be a marker for the direction of selection \cite{buffalo2019linked}.

\section{Results}

\subsection{Navigability and Ruggedness}

\begin{figure}%
    \centering    
  {\includegraphics[width=\linewidth]{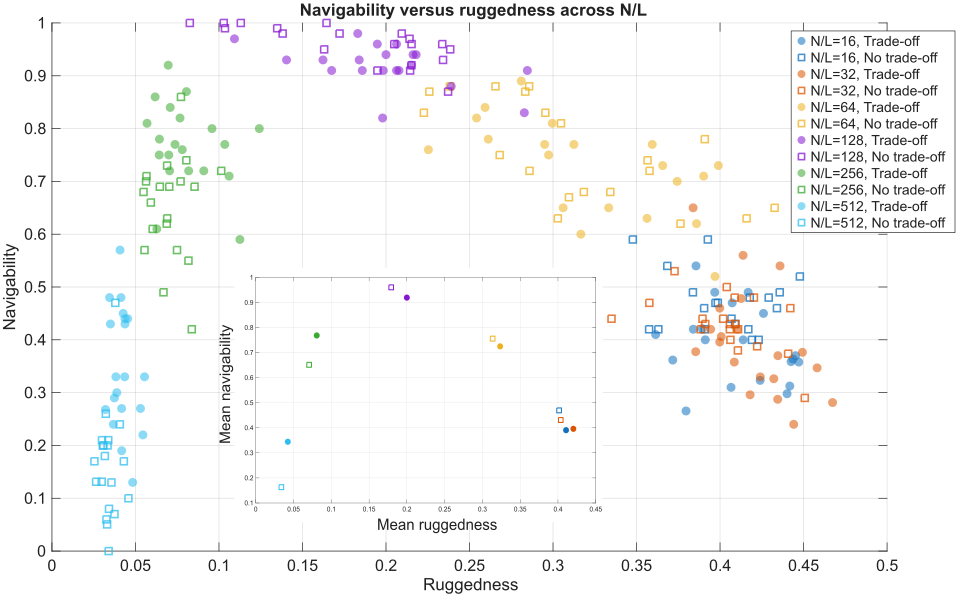}}\\%

    \caption{Navigability and ruggedness across genotype--phenotype map architectures.
Each point represents one independently generated map. Filled circles denote the trade-off fitness function, while open squares denote the no-trade-off control. Colours indicate the width-to-depth ratio \(N/L\). Navigability is highest at intermediate \(N/L\), especially around \(N/L=128\), and decreases again at very large \(N/L\), despite the continued decrease in ruggedness. Thus, low ruggedness alone is not sufficient for high navigability. The trade-off has little effect at small and intermediate \(N/L\), but improves navigability relative to the no-trade-off control at large \(N/L\), especially for \(N/L=256\) and \(N/L=512\). \label{fig_nav} }
\end{figure}

Figure \ref{fig_nav} shows the navigability and ruggedness of different realizations of $6$ map architectures. Each point in Figure \ref{fig_nav} represents one independently generated map, while the corresponding mean plot in the inset summarizes the average behavior across maps.

The plot shows that the relationship between ruggedness and navigability is strongly architecture-dependent. For small maps with small width-to-depth ratios, \(N/L=16\) and \(N/L=32\), the landscapes are relatively rugged, and navigability is low. In this regime, the trade-off and no-trade-off landscapes largely overlap, and the trade-off does not substantially improve access to high-fitness regions. Mutational effects in the relatively short genotype in these maps are repeatedly transformed through many layers, the imposed phenotypic trade-off has little effect on the global accessibility of adaptive paths.

As \(N/L\) is increased to intermediate values, navigability rises. The highest navigability occurs at \(N/L=128\) or $N=1024$, $L=32$, where both trade-off and no-trade-off landscapes have relatively low ruggedness and high path success. In this regime, most maps have navigability close to one, indicating that low-fitness genotypes can largely reach high-fitness target regions through strictly fitness-increasing mutational paths. This suggests that intermediate map architectures produce landscapes in which local uphill moves are well connected into global adaptive routes.

However, the dependence on \(N/L\) is not monotonic. At \(N/L=256\), with a decrease $L$ the ruggedness decreases further, but navigability also decreases relative to \(N/L=128\). At \(N/L=512\), keeping $N$ the same and reducing $L$ ruggedness is lowest, but navigability falls substantially. Thus, landscapes with very few local maxima are not necessarily highly navigable. In other words, ruggedness and navigability are not simply inversely related. For wide and shallow maps, landscape can have few local traps and still lack uphill routes from low-fitness genotypes to high-fitness target regions.

The effect of the trade-off also changes with architecture. For \(N/L=16,32,64,\) and \(128\), the no-trade-off control is comparable to, and in some cases slightly more navigable than, the trade-off landscape. In contrast, at larger \(N/L\), especially \(N/L=256\) and \(N/L=512\), the trade-off landscape becomes more navigable than the no-trade-off control. At \(N/L=512\), both landscapes have very low ruggedness, but the trade-off landscape has substantially higher navigability than the no-trade-off landscape. Thus, the beneficial effect of a trade-off is not universal; it emerges only in sufficiently wide and shallow maps. This is seen most clearly by comparing $N/L=256$ and $N/L=512$ where decreasing the number of layers while keeping the genotype length fixed leads to a decrease in navigability, and a stronger effect of the trade-off on navigability. Previous work has shown how the high dimension of genotypic space can allow the evolutionary search to circumvent fitness valleys and make the landscape more navigable \cite{greenbury2022structure, papkou2023rugged}. Here we find that lowering the number of transformations between genotype and phenotype could also improve navigability.

Our results suggest that for large maps, trade-offs can help increase navigability. The strongest distinction between trade-off and no-trade-off fitness occurs where the map is sufficiently wide that phenotypic correlations can organize the accessible path structure. Evolutionary navigability is thus a path-level property of the genotype-phenotype map, rather than a direct consequence of local ruggedness alone. 

\begin{figure}
    \centering
    \includegraphics[width=1\linewidth]{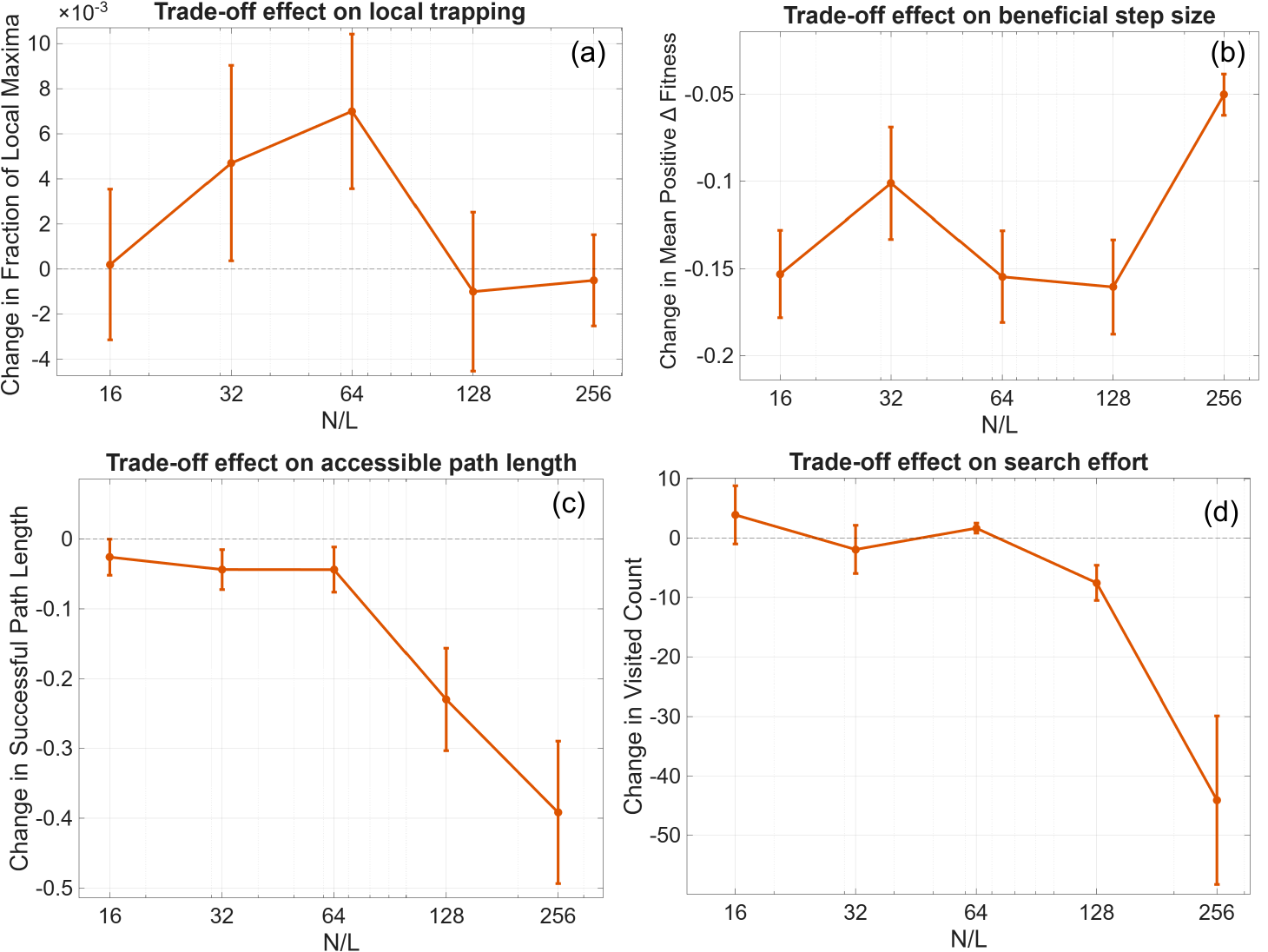}
    \caption{For each width-to-depth ratio \(N/L\), diagnostics were computed
under the trade-off fitness function and the no-trade-off control, and the plotted quantity is the paired difference \(\Delta X = X_{\text{TO}} - X_{\text{NTO}}\). Error bars denote the standard error across independently generated maps. The trade-off has little systematic effect on the fraction of local maxima (a), indicating that it does not primarily act by removing local traps. Instead, the main local effect is a reduction in the mean size of beneficial mutations, $⟨\Delta F\mid \Delta F>0⟩$ (b), consistent with partial cancellation between opposing fitness components. Despite this reduction in beneficial step size, successful accessible paths become shorter at larger $N/L$ (c), and the number of genotypes visited during search decreases strongly at high $N/L$ (d). Together, these diagnostics suggest that trade-offs reshape the connectivity and geometry of accessible paths rather than solely smoothing the landscape or increasing the number of uphill moves.}
    \label{fig:diag}
\end{figure}

In fact, the fraction of local maxima shows little systematic change under the trade-off (Figure \ref{fig:diag} (a)). Across architectures, the difference remains
close to zero, with error bars overlapping zero for most values of \(N/L\). This
indicates that the trade-off does not primarily alter navigability by removing
local traps. In other words, the improved accessibility observed at larger
\(N/L\) cannot be explained simply by a reduction in the number of local maxima.

Similarly, only a slight local effect of the trade-off is seen in the size of beneficial mutations (Figure \ref{fig:diag}(b)). The mean positive fitness
effect, is consistently smaller under the trade-off than under the no-trade-off
control. This is consistent with the trade-off fitness combining two partially
opposing phenotypic objectives. A mutation that increases one component of
fitness may decrease the other, so the net beneficial effect under
\(F_{\mathrm{TO}}\) is muted. Thus, the trade-off reduces the typical size of
available uphill steps.

Despite this reduction in beneficial step size, the path-level quantities clearly show that trade-off landscapes can become easier to traverse at larger \(N/L\).
Among successful searches, the accessible path length decreases under the
trade-off, with the effect becoming stronger for larger width-to-depth ratios
(Figure \ref{fig:diag}(c)). Since breadth-first search returns the shortest
strictly uphill path within the explored mutable subspace, this indicates that
successful trade-off paths are more direct in these architectures.

The search-effort shows the same architecture-dependent trend (Figure \ref{fig:diag}(d)). At small and intermediate \(N/L\), the number of visited genotypes is similar under the two fitness definitions. At larger \(N/L\), however, the trade-off substantially reduces the number of genotypes visited during the search. Thus, when an accessible path exists, the search requires less exploration under the trade-off in wide and shallow maps.

Together, Figure \ref{fig:diag} shows that the trade-off does not improve navigability by simply smoothing the local landscape or increasing the number of beneficial mutations. Instead, it reduces the size of individual beneficial steps while reorganizing how those steps connect into larger uphill paths. This explains why ruggedness and local trapping alone do not predict navigability; the key effect of the trade-off lies in the connectivity of accessible paths through the genotype-phenotype map.

\subsection{Epistasis \label{sec_epi}}

\begin{figure}%
    \centering    
  {\includegraphics[width=\linewidth]{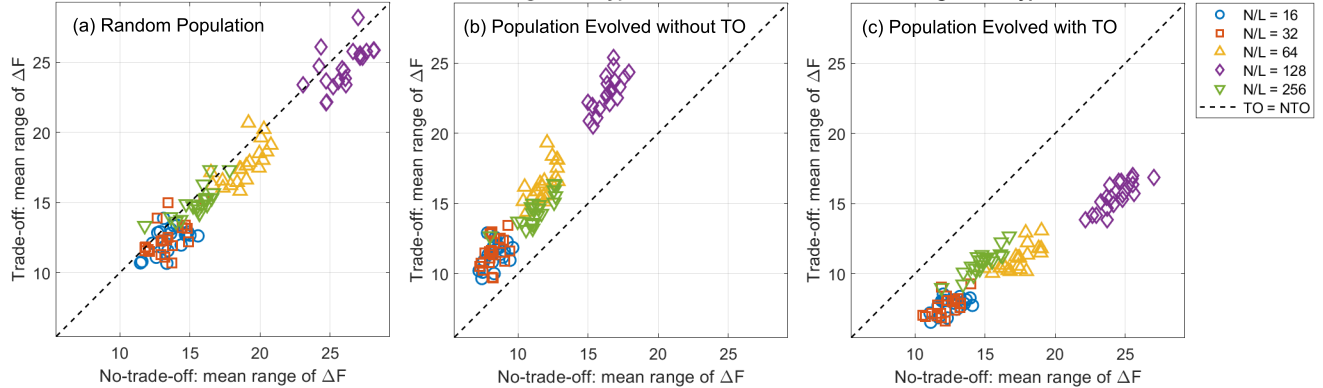}}\\%

    \caption{Each point represents one independently generated map, with colors and markers denoting the architecture $N/L$. The x-axis shows the mean range of single-mutant fitness effects, $\Delta F$, when fitness is evaluated under the no-trade-off objective, while the y-axis shows the same quantity under the trade-off objective. The dashed line marks equality between the two fitness definitions. In random populations (a), points lie close to the diagonal, indicating that the trade-off does not strongly alter mutation-effect range in unconditioned genotype space. In populations evolved without the trade-off (b), points lie mostly above the diagonal, showing that the trade-off objective would produce a larger range of mutational effects on those genetic backgrounds. In contrast, in populations evolved with the trade-off (c), points fall below the diagonal, indicating that trade-off evolution leads populations into regions where the trade-off fitness has a reduced range of mutational effects. Thus, the trade-off does not globally change epistasis across all backgrounds; rather, it changes which regions of the genotype–phenotype map are sampled by evolution.\label{fig_epistasis} }
\end{figure}

\begin{figure}%
    \centering    
  {\includegraphics[width=\linewidth]{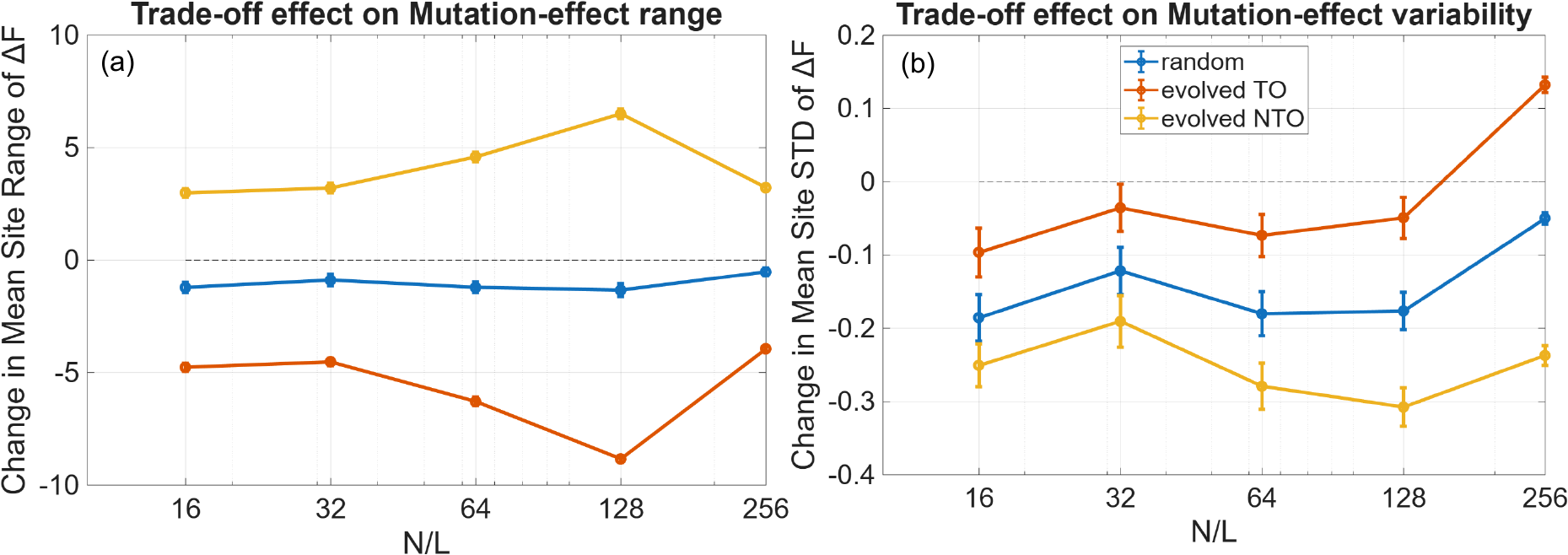}}\\%

    \caption{For each architecture, we plot the paired difference between the trade-off and no-trade-off fitness definitions, $X_{TO}-X_{NTO}$, for three genotype backgrounds: random populations, populations evolved under the trade-off objective, and populations evolved under the no-trade-off objective. The dashed horizontal line marks no difference between the two fitness definitions. (a) Change in the mean range of single-mutant fitness effects, $\Delta F$, across genetic backgrounds. Populations evolved under the trade-off show a reduced mutational-effect range when evaluated with the trade-off objective, whereas populations evolved without the trade-off show the opposite pattern, indicating that evolution locally buffers the objective under which the population was selected. (b) Change in the mean site-wise standard deviation of $\Delta F$, measuring typical background-dependence of mutational effects. Unlike the range, this measure shows a more modest and mostly negative trade-off effect across backgrounds, indicating that typical and extreme background-dependence need not vary in the same way. Together, these results suggest that trade-offs do not globally alter epistasis in a uniform manner; rather, their effect depends on evolutionary history and GP-map architecture.\label{fig_epistasis2} }
    
\end{figure}

Figure \ref{fig_epistasis} shows the trade-off and no trade-off range of fitness effects of a single mutation for a randomly chosen population, and populations evolved with and without a trade-off for our different network architectures. It is interesting to note that across the different populations, the architecture with $N/L=128$ shows the highest range of fitness effects across sites. Thus, the network architecture and complexity biases the system towards or away from epistasis in a significant way.

The plots also show that trade-offs do not globally alter epistasis in a uniform way. In random genetic backgrounds, the trade-off and no-trade-off objectives produced similar patterns of mutation-effect background-dependence, suggesting that much of this structure is imposed by the architecture of the genotype--phenotype map itself. However, in evolved populations the comparison became history-dependent. Populations evolved under the trade-off objective tended to occupy regions where the trade-off fitness had a reduced range of mutational effects, whereas populations evolved under the no-trade-off objective tended to occupy regions where the no-trade-off fitness had a reduced range of mutational effects.

Thus, the epistasis sweep supports the interpretation that the trade-off does not simply increase or decrease epistasis everywhere in genotype space. Rather, the multilayer genotype--phenotype map creates an epistatic substrate, and evolutionary history determines which regions of this substrate are sampled. Trade-offs therefore reshape the local mutational neighborhoods encountered by evolving populations, rather than globally rewriting the background-dependence of mutation effects across all genotypes.

 Figure \ref{fig_epistasis2} shows the range and standard-deviation diagnostics capture distinct aspects of background-dependence. The range is sensitive to rare backgrounds in which a mutation has an unusually large positive or negative effect, whereas the standard deviation measures the typical spread of mutation effects across backgrounds. Their different trends suggest that trade-offs and evolutionary history affect the tails and the bulk of the mutation-effect distribution differently.

For evolved TO populations, the range plot shows that the trade-off has a smaller extreme spread than NTO. But the standard deviation plot says the typical spread is not reduced as strongly, and at N/L=256 may even increase. So TO-evolved populations may be eliminating extreme trade-off sensitive backgrounds while still retaining moderate variation in ordinary mutation effects.

For evolved NTO populations, the range plot says trade-offs have a much larger extreme spread than NTO, but the standard deviaton plot says typical variability is lower under trade-offs. That suggests the trade-off fitness on NTO-evolved backgrounds may have rare extreme responses, even if most mutation effects are not broadly more variable.

Evolution under an objective (either with trade-off or without) reduces the extreme mutational range of that objective, but typical mutation-effect variability and extreme mutation-effect variability need not change together.

\subsubsection{Microevolution in a High Fitness Population}
\begin{figure}%
    \centering    
  {\includegraphics[width=\linewidth]{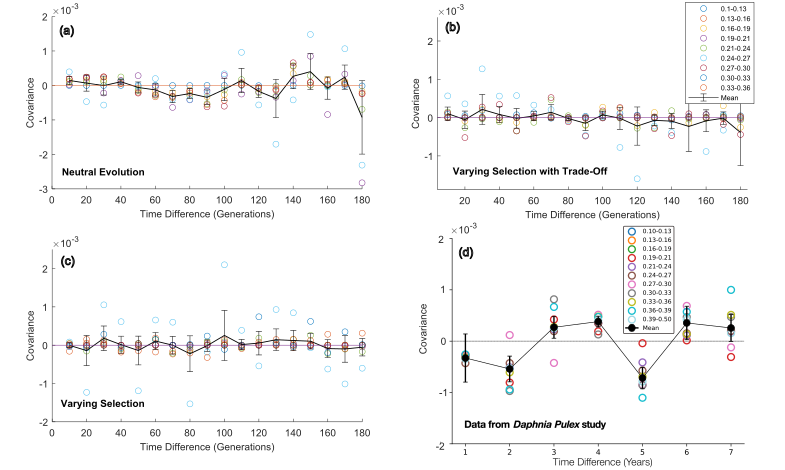}}\\%
    \caption{Covariance over time of minor allele frequency changes for a high fitness population under (a) neutral evolution, (b) varying selection that presents a trade-off and (c) varying selection without a trade-off. Figure (d) shows the covariance of minor allele frequency changes calculated from data obtained from \cite{lynch2024genome}. We find high fitness genotypes and allow them to evolve under selection from the same fitness function. We then look at minor allele frequencies and calculate the covariance in changes in these frequencies over different time gaps. The covariance values for the first points have the mean of $19$ different time difference intervals, while the last only has one time difference interval. In (b) and (c), the covariance is calculated in a manner parallel to (a), but with the environment switching between two different ones every $10$ generations. The differences between the three cases are slight. The covariance for varying selection with and without a trade-off remain near zero, with both positive and negative fluctuations. The covariance for neutral evolution shows more sustained shifting with the covariance remaining negative or positive for more than one time difference. In (d) the covariance also shows changes in direction, with the value remaining near zero. Note that the time scale on figures (a)-(c) are arbitrary since they depend on choice of mutation rate and the simulated population's growth rate.  \label{fig_covariance}}
\end{figure}

Figure \ref{fig_covariance} (a) shows the covariance over time of minor allele frequency changes for neutral evolution. Similar to the results from \cite{lynch2024genome}, shown in Figure \ref{fig_covariance} (d) we see a variation in the sign of covariance, although in our simulation alternating selection is absent. Further, Figure \ref{fig_covariance} (b) and (c) shows the covariance over time of minor allele frequency changes for varying selection pressures, with and without a trade-off. Although these look broadly similar to the case with neutral selection, some subtle differences are present. Both cases with alternating selection show slightly more rapid changes in sign than the neutral case. The neutral case shows covariance that is slower to switch signs. Further, the covariance remains smaller in absolute magnitude for the case with alternating selection. Note that the time scale on Figure \ref{fig_covariance} (a)-(c) is arbitrarily chosen since generation time is dependent upon mutation rate and growth rate of the simulated population. We are aiming for a qualitative comparison with the study in \cite{lynch2024genome}.

However, the fact that we see alternating covariance in minor allele frequency changes even in the absence of varying selection is significant. We believe that the presence of an alternating covariance could point to the population switching between competing high fitness genotypes that already exist in the population, rather than external changes in the environmental selection. 

We posit that the switching in covariance of changes in minor allele frequencies is due to the presence of epistasis in the system. The fitness landscape is rugged, with several high fitness peaks \cite{whitlock1995multiple}. As the population switches from one high fitness genotype to another, it is possible that mutations to minor alleles which were previously deleterious are now beneficial due to epistasis and vice versa. In other words, it is possible that the fitness effect of mutations in minor alleles depends on the genetic background.

\section{Discussion}

Classical NK-type models established how epistatic interactions can tune landscape ruggedness by imposing ruggedness directly at the genotype--fitness level rather than through an explicit developmental or genotype--phenotype architecture \cite{kauffman1987towards, kauffman1989nk}. Work on holey adaptive landscapes, neutral networks, and genotype networks then showed that high-dimensional genotype spaces can contain connected sets of viable or phenotypically similar genotypes, making rugged landscapes more navigable than a low-dimensional peaks and valleys picture would suggest \cite{gavrilets1997evolution, gavrilets1997percolation, schuster1994sequences, wagner2008robustness, weinreich2006darwinian}. More recent work has made this point explicit by showing that structural properties of genotype--phenotype maps can make fitness landscapes navigable \cite{greenbury2022structure, papkou2023rugged}. Separately, multilevel genotype--phenotype models such as toyLIFE have shown that adding levels of phenotypic organization can change robustness and evolvability \cite{arias2014toylife, catalan2018adding}.

Our work combines these strands but adds a missing ingredient; an explicit, tunable multilayer genotype--phenotype map together with an implemented fitness trade-off. We ask how navigability, ruggedness and epistasis depends on the architecture of the map; its width, depth, and width-to-depth ratio, and on whether fitness is defined by compatible or conflicting phenotypic objectives.  The trade-offs we consider are not ad-hoc but emerge from the structure of the map, and are brought into evolutionary considerations through our construction of fitness. Our work contributes to relatively few studies that consider the joint effect of genotypic-phenotype map complexity and environmental constraints.  

We explicitly model how environmental constraints enter through the mapping from phenotype to scalar fitness. In sequence-to-structure GP maps, molecular structures such as RNA folds or protein conformations are treated as phenotypes \cite{schuster1994sequences, ferrada2012comparison, ahnert2017structural}. These are phenotypes on the molecular scale, but are usually intermediate traits from the perspective of organism fitness. In such models, the downstream mapping from molecular structure to fitness is typically collapsed into a direct phenotype--fitness assignment. In contrast, our multilayer model represents additional stages of genotype-phenotype mediation before fitness is evaluated. We investigate different architectures of the map, with different $N$ and $L$ values. 

Sequence-to-structure GP-map studies can represent different environments by changing the fitness assigned to each phenotype \cite{greenbury2022structure, chevin2022using}. In our model, the environment enters through competing phenotypic objectives imposed after a multilayer genotype--phenotype map, allowing us to study how map architecture and environmental trade-offs jointly shape navigability and epistasis. Crucially, fitness can be evaluated either with or without a phenotypic trade-off. In the trade-off condition, fitness combines two negatively correlated phenotypic components, whereas the no-trade-off control combines one of these components with an uncorrelated phenotypic target.

Our results in Figure \ref{fig_nav} connect to the broader view that ruggedness does not necessarily preclude evolutionary accessibility in high-dimensional genotype spaces. In the classical low-dimensional picture of a fitness landscape, ruggedness is often identified with evolutionary constraint since local maxima trap populations and prevent further adaptation. However, genotype spaces are high-dimensional graphs, and genotype--phenotype maps can contain large connected sets of genotypes with identical or similar phenotypes. As emphasized in previous work, these genotype networks can make rugged landscapes navigable by providing many alternative routes through sequence space \cite{wagner2008robustness,papkou2023rugged, ciliberti2007innovation}.

Building on this idea, we find that navigability is not simply the correlated with ruggedness. In particular, the \(N/L=512\) maps have low ruggedness but also low navigability, showing that a scarcity of local maxima is not sufficient for reliable access to high-fitness regions. For these maps $N=4096$ and $L=8$. When compared to maps with $N=4096$ and $L=16$ or $L=32$, we find a decrease in navigability  with decrease in number of layers despite the genotype remaining high dimensional. This suggests that the depth of the map plays a crucial role in determining navigability and a high dimensional genotype does not guarantee high evolvability. Conversely, navigability is highest at intermediate architectures, especially around \(N/L=128\), where local uphill moves appear to connect efficiently into global adaptive routes. Thus, what matters is not only the number of local traps, but the connectivity of the directed graph of fitness increasing mutations.

The effect of trade-offs further shows that this connectivity depends on how phenotypic variation is projected onto fitness. Surprisingly, trade-offs work to increase navigability across some architectures, although not uniformly across architectures. Rather, its effect emerges most clearly in maps with high $N/L$ values, where the negatively correlated phenotypic components can organize accessible paths into smoother routes. This suggests that trade-offs can alter the global topology of adaptive paths without simply smoothing the local landscape. In this sense, our model extends the genotype-network view of navigability by showing that path accessibility depends jointly on high-dimensional genotype-space structure, genotype-phenotype map architecture, and the multi-objective structure of fitness.

Figure \ref{fig:diag} shows that beneficial mutations are typically smaller under the trade-off, consistent with partial cancellation between opposing phenotypic components. Nevertheless, in wide and shallow maps the trade-off reduces successful path length and search effort. This indicates that the trade-off changes the geometry of the accessible graph: it can make uphill moves connect more directly into successful routes even while reducing the size of individual fitness gains. 

Trade-offs in evolutionary objectives can arise from epistatic interactions as well as environmental variation. Our model offers a way to study the complex interaction between epistasis and environmental background \cite{hall2019environment}. Figure \ref{fig_epistasis} suggests that trade-offs do not globally increase or decrease background-dependence. In random genotypes, the trade-off and no-trade-off objectives give similar mutation-effect ranges, indicating that much of the epistatic structure is imposed by the GP-map architecture itself. However, after evolution, the comparison becomes history-dependent. Populations evolved under a given objective occupy regions in which that objective is locally buffered against mutation. Thus, the apparent effect of trade-offs on epistasis depends on the genetic background produced by prior evolution, consistent with previous experimental studies \cite{hall2019environment}.  However, intermediate architectures with $N/L=128$ show higher degree of epistasis across evolutionary paradigms. Thus, the map architecture creates the substrate of background-dependent mutation effects, while evolutionary history determines which parts of that substrate are sampled. 

The difference between the range and standard-deviation of single mutation fitness effects in Figure \ref{fig_epistasis2} shows that background-dependence is not captured by a single scalar measure. The range is sensitive to rare backgrounds in which a mutation has an unusually large positive or negative effect, whereas the standard deviation captures typical variation across backgrounds. Their different trends suggest that trade-offs and evolutionary history affect the tails and the bulk of the mutation-effect distribution differently.

Lastly, we use our model to test a prediction for the micro-evolution of high-fitness populations. Our simulations suggest that temporal covariance patterns in allele-frequency changes seen in a recent long term study in \textit{Daphnia Pulex} can arise even in the absence of externally alternating selection \cite{lynch2024genome}. In our model, such covariance fluctuations may reflect movement among multiple high-fitness genotypes in an epistatic landscape. This does not rule out fluctuating selection, as was suggested in the experimental study, but it suggests that similar genomic signatures may also be produced by the internal structure of the genotype-phenotype map. The construction of our model allows us to test these three cases; neutral selection, alternating selection with and without a trade-off explicitly to look at the hypothesis in \cite{lynch2024genome} more closely.

A limitation of the present model is that the genotype--phenotype map itself is held fixed over the evolutionary searches. Thus, evolution acts on genotypes within a given map, but does not alter the architecture of the map through changes in regulatory interactions, developmental pathways, or network topology. This separates the question of how a fixed genotype--phenotype architecture shapes accessible evolutionary paths from the broader question of how such architectures themselves evolve over longer timescales \cite{davidson2006gene, ciliberti2007innovation, manrubia2021genotypes}.

Our map is also strictly feed-forward. Mutational effects are propagated from the genotype layer to the phenotype layer through a sequence of directed transformations, with no feedback from later layers to earlier ones and no recurrent dynamics within a layer. However, biological regulatory and developmental systems often contain feedback loops, recurrent regulatory interactions, and dynamical attractors that stabilize cell states and shape developmental trajectories
\cite{davidson2003regulatory,alon2007network, jaeger2014bioattractors}. Incorporating such feedback would be an important extension, because recurrent network dynamics could change how mutational effects are buffered, amplified, or canalized before they reach fitness.

\section{Data}
The simulation code used to obtain our central results can be found at \url{https://github.com/nanditacha/GPmap}.

\bibliographystyle{unsrt}
\bibliography{biblio}
\end{document}